\documentclass[twocolumn,floatfix,aps,prd,showpacs,amsmath,amssymb,nofootinbib,preprintnumbers,superscriptaddress,natbib]{revtex4}

\usepackage{graphicx}

\usepackage{array}

\begin{document}

\title{Forecast Constraints on Inflation from Combined CMB and Gravitational Wave Direct Detection Experiments}

\author{Sachiko Kuroyanagi} \email[]{s-kuro@a.phys.nagoya-u.ac.jp}
\affiliation{Department of Physics, Nagoya University, Chikusa, Nagoya
  464-8602, Japan}

\author{Christopher Gordon} 
\affiliation{Astrophysics Department, University of Oxford, Denys Wilkinson Building, Keble Road, Oxford OX1 3RH, UK}

\author{Joseph Silk} 
\affiliation{Astrophysics Department, University of Oxford, Denys
Wilkinson Building, Keble Road, Oxford OX1 3RH, UK} \affiliation{Institut d' Astrophysique, 98bis Boulevard Arago, Paris 75014, France}

\author{Naoshi Sugiyama} \affiliation{Department of Physics, Nagoya
  University, Chikusa, Nagoya 464-8602, Japan} \affiliation{Institute
  for Physics and Mathematics of the Universe, University of Tokyo,
  5-1-5 Kashiwa-no-ha, Kashiwa City, Chiba 277-8582, Japan}

\begin{abstract}
We study how direct detection of the inflationary gravitational wave
background constrains inflationary parameters and compliments
CMB polarization measurements.  The error ellipsoids calculated using
the Fisher information matrix approach with Planck and the direct
detection experiment, Big Bang Observer (BBO), show different directions
of parameter degeneracy, and the degeneracy is broken when they are
combined.  For a slow-roll parameterization, we show that BBO could
significantly improve the constraints on the tensor-to-scalar ratio
compared with Planck alone. We also look at a quadratic and a natural
inflation model.  In both cases, if the temperature of reheating is also
treated as a free parameter, then the addition of BBO can significantly
improve the error bars.  In the case of natural inflation, we find that
the addition of BBO could even partially improve the error bars of a
cosmic variance-limited CMB experiment.
\end{abstract}


\maketitle

\section{Introduction}
Inflation \cite{inf1,inf2,inf3}, which is widely believed to have taken
place in the very early universe, generically predicts the existence of
tensor mode perturbations originating from quantum fluctuation
\cite{grav1,grav2,grav3}.  Since inflation predicts an almost
scale-invariant spectrum, the tensor mode perturbations are considered
to exist as a gravitational wave background over a wide range of
frequencies.  Because of their weakness of interaction with matter and their
linearity, they are expected to remain uncontaminated even at higher
frequencies, whereas the scalar mode perturbations become nonlinear and
it might seem difficult to recover information about its spectrum on
smaller scales.  Therefore the gravitational wave background possesses
the potential to enable us to reconstruct the primordial spectrum over a
wider range of scales. This helps us learn more about the inflation
potential since the amplitude of the spectrum at each scale corresponds
to the height of the potential when the mode exited the horizon during
inflation.

One of the ways to detect the inflationary gravitational wave background
is to measure the $B$-mode polarization of the cosmic microwave
background (CMB) induced by primordial gravitational waves
\cite{CMBp1,CMBp2}.  The ongoing satellite mission, Planck
\cite{Planck}, is expected to detect the signature of the large-scale
tensor perturbations if they are sufficiently large in amplitude.  The
other way of probing the gravitational wave background is to detect it
directly with space-based laser interferometers such as Big Bang
Observer (BBO) and DECi-hertz Interferometer Gravitational wave
Observatory (DECIGO) \cite{bbo,decigo}.  Clearly, both approaches
provide us with new information about the early universe, and help in
determining inflationary parameters more accurately if they detect the
gravitational wave background.  It is notable that while CMB $B$-mode
experiments measure the gravitational wave background at the present
horizon scale ($\sim 10^{-18}$Hz), direct detection experiments measure
it at much smaller scales which correspond to the detector size ($\sim
0.1-1$Hz).  This means that these two different types of experiments
could provide independent information.

In this paper, our goal is to predict how the direct detection of the
gravitational wave background can complement CMB observations.  The
complementarity of the two observations is discussed within the slow-roll
paradigm in Refs. \cite{Smith1, Smith2, Ungarelli}.  While they only
discuss the detectability of the gravitational wave background in 
direct detection experiments by connecting the amplitude of the
gravitational wave background at direct detection scales with model
parameters which are allowed by the current CMB constraints, we perform
the Fisher matrix calculation and forecast errors on the parameters
attainable in the direct detection experiments.  This enables us to
discuss how well inflation parameters may be determined by direct
detection of the gravitational wave background and how degeneracy in
parameter space may be broken when the two constraints from CMB
observation and direct detection are combined.

Here, we consider the combination of the upcoming CMB satellite
experiment Planck and the future satellite gravitational wave detector
BBO.  As long as we make the usual assumption that the equation of state
is $\ge-1$ during inflation, the spectrum of the gravitational wave
background cannot be blue-tilted and sensitivities of ground-based
experiments and the preceding space mission LISA are not enough to
detect the inflationary gravitational waves.  Therefore, it would be a
long road to seek the inflationary gravitational waves, while CMB
experiments are expected to detect $B$-mode signals in the not so distant
future.  One may think constraints on inflationary parameters may be
improved by post-Planck CMB experiments before direct detection is
achieved by BBO.  For this reason, we also look at the case combined
with a cosmic variance-limited experiment, which would be similar to CMB
satellite missions planned for a few decades like BBO, instead of
Planck.

Applying the Fisher matrix method to the specification of Planck and
BBO, we evaluate errors on inflationary parameters expected in these two
future experiments.  To calculate the Fisher matrices, we need to obtain
differentiation of the spectrum with respect to the model parameters
assuming a fiducial model.  The CMB spectra are computed using the CAMB
code \cite{Lewis}.  For the gravitational wave spectra at direct
detection scales, we present two different approaches: an analytical way
with the slow-roll approximation and a numerical way.  We first apply
the Fisher matrix method to the slow-roll paradigm, which enables us to
express the spectrum in terms of the parameters defined at the CMB
scale.  This slow-roll framework is a simple and well-established way to
connect the amplitude of the spectrum at different scales.  However,
since this method uses the Taylor expansion to extend the spectrum from
the CMB scale, the approximated spectrum may deviate slightly from the
true value at the direct detection scale \cite{Kuroyanagi}.  For this
reason, we also present the case in which numerical calculations are
used to obtain the precise amplitude of the spectrum.  This method,
however, requires us to assume an inflation model.  Therefore, both of
the methods have their advantages and disadvantages; the slow-roll
paradigm can be applied as a more general model, which does not require
us to assume an inflation model like the numerical method, while the
numerical calculation enables us to make a more accurate prediction.

The outline of this paper is as follows: In Sec. \ref{Fishermethod},
we briefly introduce the Fisher matrix methods for both CMB measurements
and direct detection of the gravitational wave background.  In
Sec. \ref{slowroll}, the Fisher method is applied to the slow-roll
paradigm.  We show how direct detection reduces errors obtained from CMB
in the $n_S-r$ plane by connecting $n_S$ and $r$ with parameters of the
gravitational wave spectrum assuming slow-roll inflation.  In
Sec. \ref{numerical}, the investigation with numerically calculated
spectra is given for several inflation models which predict a large
enough amplitude of the gravitational waves to be detected by BBO.  We
also allow the temperature of reheating to be a free parameter.  We show
to what extent potential parameters can be determined when the
constraints from CMB and direct detection are combined.  Conclusions
are given in Sec. \ref{conclusion}.

\section{Fisher matrix methods}
\label{Fishermethod}
The Fisher information matrix is commonly used in cosmology to predict
how well parameters can be determined in future planned experiments,
and is defined as the second derivative of the log likelihood function
${\cal L}$ with respect to the model parameters $p_i$,
\begin{equation}
{\cal F}_{ij}=-\left\langle\frac{\partial^2 \ln{\cal L}}
{\partial p_i \partial p_j}\right\rangle.
\end{equation}  
According to the Cramer-Rao inequality, the Fisher matrix gives a lower
bound for variances of the parameter estimates.  This enables us to estimate
expected errors on model parameters for a given experiment.  The
likelihood function is constructed from the specifications of each
experiment.

\subsection{CMB $B$-mode polarization}
\label{CMBFisher}
The CMB  Fisher matrix is given by (see for example \citep{zalspesel97})
\begin{equation}
{\cal F}_{ij}=\sum_{\ell} \sum_{X,X'} \frac{\partial C_{\ell}^X}{\partial p_i} {\rm Cov}^{-1} (C_{\ell}^X,C_{\ell}^{X'})\frac{\partial C_{\ell}^{X'}}{\partial p_j},
\end{equation}
where $X$ and $X'$ are summed over the the CMB temperature, $E$-mode, and
$B$-mode of the CMB polarization.  The covariance matrix can be obtained
from Zaldarriaga et. al. \cite{zalspesel97} and  depends on the
temperature noise per pixel ($\sigma_T$), the polarization noise per
pixel ($\sigma_E$ and $\sigma_B$), the pixel area in radians squared
($\theta^2=4\pi/N_{\rm pix}$), and the beam window function which we
approximate as Gaussian (${\cal B}_\ell\approx
\exp(-\ell(\ell+1)\sigma_b^2$). The values we use are taken from the
Planck blue
book\footnote{http://www.rssd.esa.int/SA/PLANCK/docs/Bluebook-ESA-SCI(2005)1\_V2.pdf}
and are listed in Table~\ref{planckparam} (note that $\theta$ needs to
be converted to radians).
\begin{table}\centering
\caption{\label{planckparam} Planck instrument characteristics \label{planckparm}}
\begin{tabular}{lcccc}
\hline
Center frequency (GHz) &70&100&143&217\\ \hline
$\theta$ (FWHM arcmin) &14&10&7.1&5.0\\
$\sigma_T$ ($\mu{\rm K}$)& 12.8& 6.8& 6.0& 13.1\\
$\sigma_E$ ($\mu{\rm K}$)&18.2&10.9&11.4& 26.7 \\
$\sigma_B$ ($\mu{\rm K}$)&18.2&10.9&11.4& 26.7 \\
\hline
\end{tabular}
\end{table}
We use $\sigma_b=\theta/\sqrt{8 \ln 2}$ and combine the different
frequency bands as specified in \citet{bonefsteg97}.  We take the range
in $\ell$ to be 2 to 2000. At higher $\ell$, secondary sources of
temperature and polarization will likely prohibit the extraction of
cosmological information from the primary CMB. We assume that the
foregrounds can be removed by using templates and multifrequency
information; see, for example, Efstathiou {\it et al.}
\citet{efgrapac09}.  We also consider the case of a cosmic
variance-limited (in both temperature and polarization) CMB experiment
(CV) for $\ell \leq 2000$ and without delensing.  This would be similar
to optimistic foreground removal with epic-2m \cite{Baumann}.

Throughout this paper, we assume a flat $\Lambda$ cold dark matter
Universe and we use the WMAP5 maximum likelihood values for the
nonprimordial power spectrum parameters \citep{WMAP5yr}: (baryon
density) $\Omega_b h^2=0.0227$, (CDM density) $\Omega_ch^2=0.108$,
(amplitude of curvature perturbations) $\Delta_{\zeta,{\rm
prim}}^2=2.41\times 10^{-9}$, (reionization optical depth) $\tau=0.089$,
and (the Hubble parameter) $h=0.724$.

\subsection{Direct detection of gravitational waves}
In a direct detection experiment, cross correlation analysis is a
powerful method to detect a weak stochastic gravitational wave
background, such as inflationary gravitational waves
\cite{Michelson,Christensen,Flanagan,Allen}.  The Fisher information
matrix for the cross correlation analysis is given as \cite{Seto}
\begin{eqnarray}
&&{\cal F}_{ij}=\left(\frac{3H_0^2}{10\pi^2}\right)^2 2T_{\rm obs}\nonumber\\
&&\times\sum_{(I,J)}\int^{\infty}_0df\frac{|\gamma_{IJ}(f)|^2\partial_{p_i}
\Omega_{\rm GW}(f)\partial_{p_j}\Omega_{\rm
GW}(f)}{f^6S_I(f)S_J(f)},
\label{Fisher}
\end{eqnarray}
where $f$ is the frequency of the gravitational waves, $H_0$ is the
Hubble constant, $T_{\rm obs}$ is observation time.  Here, we consider TDI
(Time-Delay Interferometry) channel output ($I=A,E,T$) which would be
adopted in the BBO project.  In this case, the noise transfer functions
$S_{A,E,T}(f)$ are given as \cite{Prince,LISA}
\begin{eqnarray}
S_A(f)=S_E(f)=8\sin^2(\hat{f}/2)[(2+\cos\hat{f})S^p_n(f)\nonumber\\
+2(3+2\cos\hat{f}+\cos(2\hat{f}))S^a_n(f)],
\end{eqnarray}
\begin{equation}
S_T(f)=2[1+2\cos\hat{f}]^2[S^p_n(f)+4\sin^2(\hat{f}/2)S^a_n(f)],
\end{equation}
where $\hat{f}=2\pi Lf$ and $L$ is the arm length of the detector which
is assumed the same for each arm.  In the case of the standard BBO
detector, the arm length is $L=5.0\times 10^4$km and noise functions are
$S^p_n=2.0\times 10^{-34}/L^2{\rm Hz}^{-1}$, $S^a_n=9.0\times
10^{-34}/(2\pi f)^4/(2L)^2{\rm Hz}^{-1}$. \footnote{We have taken
$S^p_n$ to be 4 times larger than the contribution from photon shot
noise alone, following Refs. \cite{Cutler,bbo}.}  The overlap reduction
function $\gamma_{IJ}(f)$ is calculated with information about relative
locations and orientations of detectors \cite{Cornish1,Cornish2,Corbin}.

The intensity of a stochastic gravitational wave background is
characterized by the dimensionless quantity 
\begin{equation}
\Omega_{\rm GW}\equiv\frac{1}{\rho_c}\frac{d\rho_{\rm GW}}{d\ln k},
\end{equation}
where the wavenumber relates to the frequency as $k=2\pi f$.  The
critical density of the Universe is defined as $\rho_{c}\equiv3H^2/8\pi
G$, where $H$ is defined by the scale factor $a(t)$ as $H=\dot{a}/a$,
and the energy density of the gravitational waves $\rho_{\rm GW}$ is
given by the 00 component of the stress-energy tensor.  Let us consider
the tensor perturbation in a Friedmann-Robertson-Walker metric, $ds^2
=-dt^2+a^2(t)(\delta_{ij}+h_{ij})dx^idx^j$.  It is convenient to expand
$h_{ij}$ into its Fourier components,
\begin{equation}
h_{ij}(t,\textbf{x})=\sum_{\lambda=+,\times}^{}\int\frac{d^3k}{(2\pi)^{3/2}}\epsilon_{ij}^{\lambda}
(\textbf{k})h_\textbf{k}^{\lambda}(t)e^{i\textbf{k}\cdot\textbf{x}},
\end{equation}
where the polarization tensors $\epsilon_{ij}^{+,\times}$ satisfy symmetric
and transverse-traceless conditions and are normalized as
$\sum_{i,j}^{}\epsilon_{ij}^{\lambda}(\epsilon_{ij}^{\lambda^{\prime}})^*=2\delta^{\lambda\lambda^{\prime}}$. 
Then $\rho_{\rm GW}$ is given as
\begin{eqnarray}
\rho_{\rm GW}&=&\frac{1}{64\pi G}\langle (\partial_t
 h_{ij})^2+(\frac{1}{a}\vec{\nabla} h_{ij})^2 \rangle\nonumber\\
&=&\frac{1}{32\pi G}\int\frac{d^3k}{(2\pi)^3}\frac{k^2}{a^2}2\sum_{\lambda}^{}|h_\textbf{k}^{\lambda}|^2,
\label{rhoGW}
\end{eqnarray}
which yields
\begin{equation}
\Omega_{\rm GW}=\frac{1}{12}\left(\frac{k}{aH}\right)^2\frac{k^3}{\pi^2}\sum_{\lambda}^{}|h_\textbf{k}^{\lambda}|^2.
\end{equation}
One may use the tensor power spectrum $\Delta_h^2(k)$ instead of
$\Omega_{\rm GW}$, which is defined as 
\begin{equation}
\Delta_h^2(k)\equiv\frac{d\langle h_{ij}h^{ij}\rangle}{d\ln k}=
\frac{k^3}{\pi^2}\sum_{\lambda}^{}
|h_\textbf{k}^{\lambda}|^2.
\end{equation}

\section{Slow-roll paradigm}
\label{slowroll}
First, we present a Fisher matrix calculation making use of analytical
models of slow-roll inflation.  The basic assumption here is that
inflation is driven by a slow-rolling single scalar field and the
slow-roll approximation is valid while the scalar field rolls down the
potential from the point which relates the CMB scale to the point of the
direct detection scale.

\subsection{Slow-roll prediction for the spectrum of 
the gravitational wave background}
\label{srformula}
In an analytic approach, it is convenient to divide the spectrum of the
gravitational wave background into two parts: the initial power spectrum
$\Delta_{h,{\rm prim}}^2$ and the transfer function $T_h$,
\begin{equation}
\Omega_{\rm GW}=\frac{1}{12}\left(\frac{k}{aH}\right)^2 \Delta_{h,{\rm prim}}^2(k)T_h^2(k).
\label{OmegaGW}
\end{equation}

The initial power spectrum is predicted under the slow-roll
approximation.  In standard slow-roll inflation, a scalar field $\phi$
slowly rolls down its potential $V(\phi)$, and the equation of motion is
given as $\ddot{\phi}+3H\dot{\phi}+V^{\prime}(\phi)=0$, where the dot
and prime denote the derivative with respect to $t$ and $\phi$
respectively.  We define the slow-roll parameters in terms of $V$ and
its derivatives,
\begin{eqnarray}
&&\epsilon\equiv\frac{m_{\rm Pl}^2}{16\pi}\left(\frac{V^{\prime}}{V}\right)^2,\\
&&\eta\equiv\frac{m_{\rm Pl}^2}{8\pi}\frac{V^{\prime\prime}}{V},
\end{eqnarray}
where $m_{\rm Pl}=1/\sqrt{G}$ is the Planck mass.  The initial power spectra
of scalar and tensor perturbations are predicted as 
\begin{eqnarray}
&&\Delta_{\zeta,{\rm prim}}^2(k)\simeq\frac{1}{\pi\epsilon}\left.\left(\frac{H}{m_{\rm Pl}}\right)^2\right|_{k=aH},\label{PSprim}\\ 
&&\Delta_{h,{\rm prim}}^2(k)\simeq\frac{16}{\pi}\left.\left(\frac{H}{m_{\rm Pl}}\right)^2\right|_{k=aH},\label{PTprim}
\end{eqnarray}
which give the tensor-to-scalar ratio
\begin{equation}
r\equiv\frac{\Delta_{h,{\rm prim}}^2(k)}{\Delta_{\zeta,{\rm
 prim}}^2(k)}\simeq 16\epsilon.
\label{tsratio} 
\end{equation}
One may define inflationary parameters at a pivot wavenumber $k_0$ and
express the initial power spectrum in a Taylor-expanded form as
\cite{Lidsey,Kosowsky}
\begin{equation}
\ln\frac{\Delta_{h,{\rm prim}}^2(k)}{\Delta_{h,{\rm prim}}^2(k_0)}=n_T(k_0)\ln\frac{k}{k_0}+\frac{1}{2}\alpha_T(k_0)\ln^2\frac{k}{k_0}+\cdots,
\label{srspectrum}
\end{equation}
where the tensor spectral index $n_T(k)$ and its running $\alpha_T(k)$
are given in terms of the slow-roll parameters 
\begin{eqnarray}
&&n_T(k)\equiv\frac{d\ln \Delta_{h,{\rm prim}}^2(k)}{d\ln k}\simeq-2\epsilon,\label{nT}\\
&&\alpha_T(k)\equiv\frac{dn_T(k)}{d\ln k}\simeq 4\epsilon\eta-8\epsilon^2.
\end{eqnarray}
Similarly, the scalar spectral index $n_S(k)$ and its running
$\alpha_S(k)$ are given as
\begin{eqnarray}
&&n_S(k)-1\equiv\frac{d\ln \Delta_{\zeta,{\rm prim}}^2(k)}{d\ln k}\simeq -6\epsilon+2\eta,\\
&&\alpha_S(k)\equiv\frac{d n_S(k)}{d\ln k}\simeq -16\epsilon\eta+24\epsilon^2+2\xi^2,
\end{eqnarray}
where $\xi^2\equiv(m_{\rm Pl}/2\pi)^4 V^\prime V^{\prime\prime\prime}/V^2$.
Throughout this paper, we use parameter values evaluated at the CMB
scale $k_0=0.002{\rm Mpc}^{-1}$.  From Eqs. (\ref{tsratio}) and
(\ref{nT}), we obtain a  relation of single-field slow-roll
inflation, which is called the consistency relation
\begin{equation}
r=-8n_T.
\label{consistencyrelation}
\end{equation}
From the relations of the slow-roll prediction, the tensor mode
parameters can be connected with the other parameters as
$n_T(k_0)\simeq-r/8$ and $\alpha_T(k_0)\simeq
r/8\left(n_S+r/8-1\right)$.  Therefore, in the framework of slow-roll
inflation, the primordial spectrum can be written in terms of the
parameters familiar in CMB observation, $r$, $n_S$ and $\Delta_{\zeta,{\rm
prim}}^2$, as
\begin{eqnarray}
&&\Delta_{h,{\rm prim}}^2(k)=r\Delta_{\zeta,{\rm prim}}^2\nonumber\\
&&\times\exp{\left[-\frac{r}{8}\ln\frac{k}{k_0}+\frac{r}{16}\left(n_S+\frac{r}{8}-1\right)\ln^2\frac{k}{k_0}\right]}.
\label{PTsr}
\end{eqnarray}

The transfer function for the simple case where the components of the
Universe are only radiation and matter is given in Ref. \cite{Turner} as
$T_h^2(k)=(3j_1(k\tau_0)/k\tau_0)^2(1+1.34x_{\rm eq}+2.5x_{\rm eq}^2)$, where
$x_{\rm eq}=k/k_{\rm eq}$, $k_{\rm eq}\equiv\tau_{\rm eq}^{-1}=6.22\times
10^{-2}\Omega_m h^2 {\rm Mpc}^{-1}$, and $\tau_0=2H_0^{-1}$.  The
spherical Bessel function, $j_1(x)=(\sin x-x\cos x)/x^2$, is replaced as
$j_1(k\tau_0)\rightarrow 1/(\sqrt{2}k\tau_0)$ when taking the limit of
$k\tau_0\ll 1$ and averaging the oscillation.  Additionally, the
amplitude of the spectrum is suppressed by the cosmological constant
\cite{Zhang} and changes in the effective degrees of freedom during the
radiation-dominated era \cite{Schwarz} at direct detection frequencies.
A suppression factor due to the cosmological constant is
$(1-\Omega_{\Lambda})^2$, explained in the Appendix.  A
damping factor due to the effective degrees of freedom is evaluated with
the temperature when the corresponding mode enters the horizon, $T_{\rm hc}$,
as $(g_*(T_{\rm hc})/g_{*0}))(g_{*s0}/g_{*s}(T_{\rm hc}))^{4/3}$, where
$g_{*0}=3.36$ and $g_{*s0}=3.90$ \cite{Watanabe}.  In the case of taking
into account only particles in the standard model and not including SUSY
particles or any other exotic particles, the effective degrees of
freedom which correspond to the direct detection scale are
$g_{*}(T_{\rm hc})=g_{*s}(T_{\rm hc})=106.75$.  Adding these two factors to the
transfer function of Ref. \cite{Turner}, we obtain \cite{Nakayama}
\begin{eqnarray}
T_h^2(k)=(1-\Omega_{\Lambda})^2\left(\frac{g_*(T_{\rm hc})}{g_{*0}}\right)\left(\frac{g_{*s0}}{g_{*s}(T_{\rm hc})}\right)^{4/3}\nonumber\\
\left(\frac{3}{\sqrt{2}(k\tau_0)^2}\right)^2(1+1.34x_{\rm eq}+2.5x_{\rm eq}^2).
\label{transfer}
\end{eqnarray}
We use this transfer function for the spectrum at the direct detection
scale, which is calculated by substituting Eqs. (\ref{PTsr}) and
(\ref{transfer}) into Eq. (\ref{OmegaGW}).

\subsection{Errors on inflationary parameters}
\begin{figure*}[htbp]
 \includegraphics[width=0.48\textwidth]{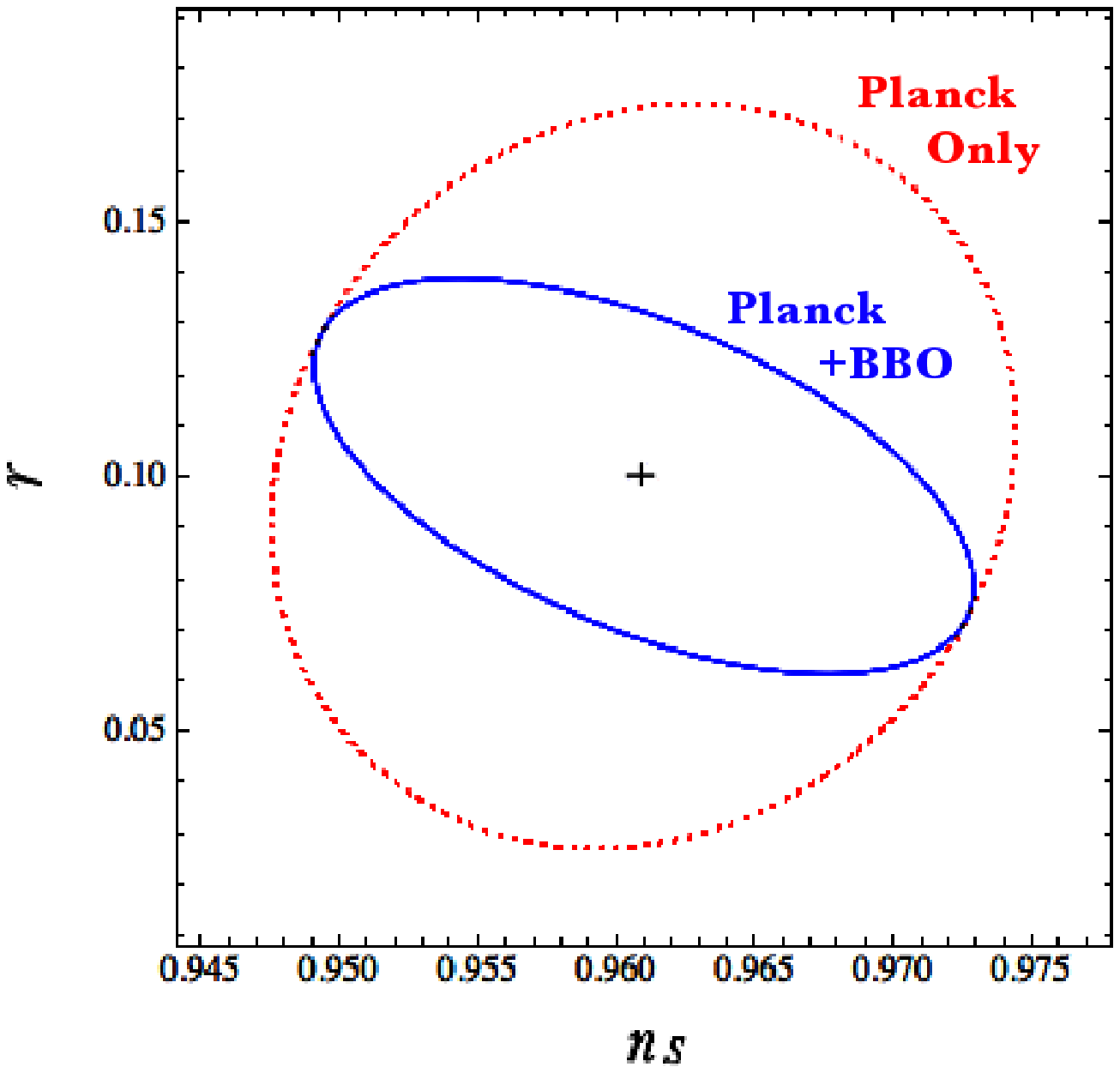} \hfill
 \includegraphics[width=0.48\textwidth]{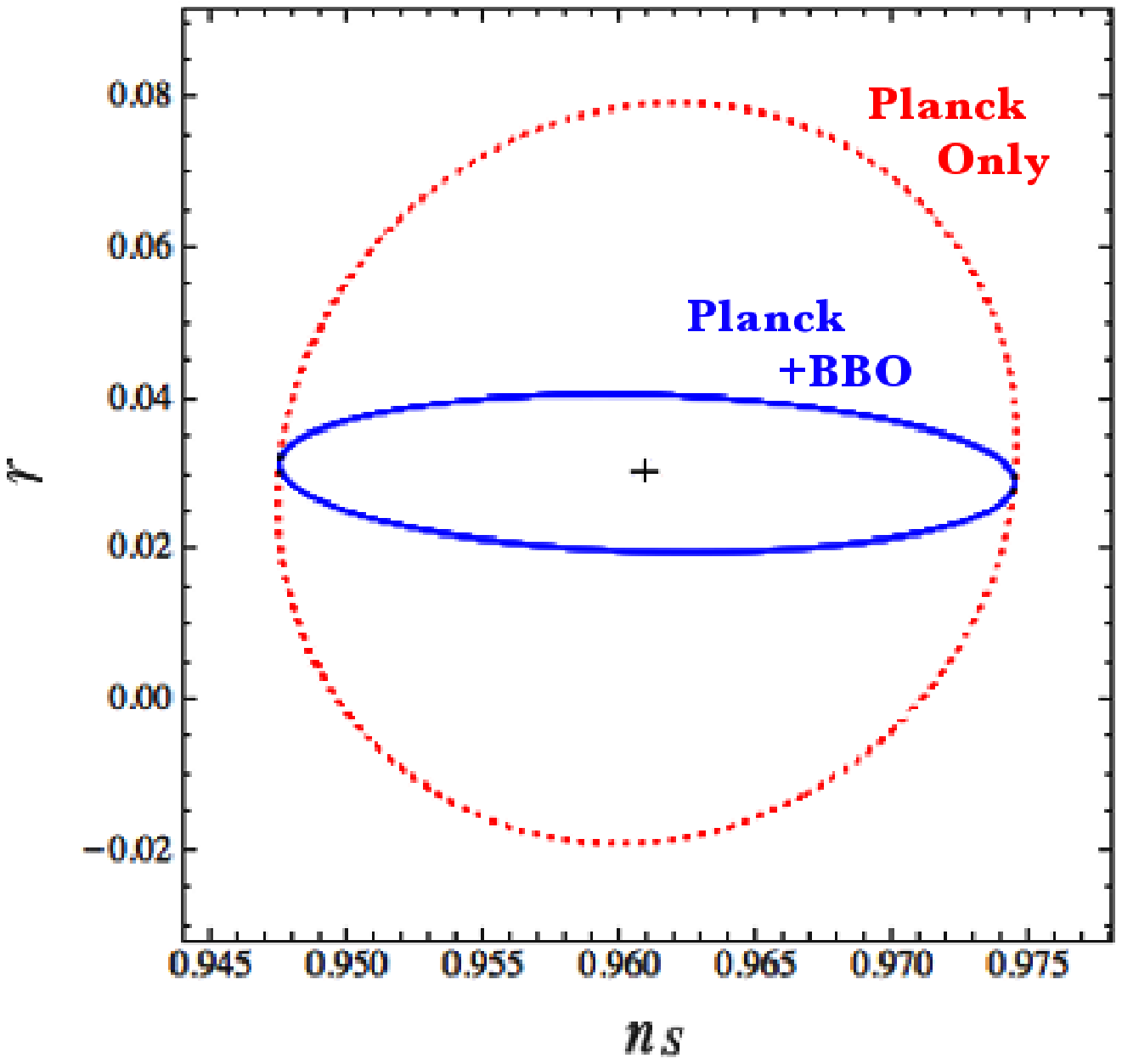} \caption{ Forecast
 constraints on $n_S$ and $r$.  The dotted line ellipse (red) represents
 the marginalized $2\sigma$ confidence level by Planck only, and
 the solid one (blue) represents combined constraints from Planck and
 BBO.  The fiducial parameters are $r_{\rm fid}=0.1$ in the left panel,
 $r_{\rm fid}=0.03$ in the right panel.}  \label{fig_sr} ss
 \makeatletter \def\@captype{table} \makeatother

\begin{equation*}
\begin{array}{c|c|cccc}
\hline
\text{Variable} & \text{Fiducial value} & \text{$\%$ Error Planck only} & \text{$\%$ Error Planck+BBO} & \text{$\%$ Error CV} & \text{$\%$ Error CV+BBO} \\
\hline
h & 0.724 & 1.1 & 1. & 0.11 & 0.11 \\
\Omega_c h^2 & 0.108 & 1.3 & 1.2 & 0.15 & 0.15 \\
\Omega_b h^2 & 0.227 & 0.88 & 0.83 & 0.13 & 0.13 \\
\tau  & 0.089 & 4.1 & 4.1 & 1.8 & 1.8 \\
n_S & 0.961 & 0.56 & 0.5 & 0.11 & 0.1 \\
r & 0.1 & 29. & 16. & 1. & 1. \\
\Delta_{\zeta,{\rm prim}}^2 & 2.41\times 10^9 & 0.79 & 0.77 & 0.29 & 0.29\\
\hline
\end{array}
\end{equation*}
\begin{equation*}
\begin{array}{c|c|cccc}
\hline
\text{Variable} & \text{Fiducial value} & \text{$\%$ Error Planck only} & \text{$\%$ Error Planck+BBO} & \text{$\%$ Error CV} & \text{$\%$ Error CV+BBO} \\
\hline
 h & 0.724 & 1.1 & 1.1 & 0.1 & 0.1 \\
 \Omega_c h^2 & 0.108 & 1.3 & 1.3 & 0.15 & 0.15 \\
 \Omega_b h^2 & 0.227 & 0.88 & 0.88 & 0.13 & 0.13 \\
 \tau  & 0.089 & 4.1 & 4. & 1.7 & 1.7 \\
 n_S & 0.961 & 0.57 & 0.57 & 0.1 & 0.1 \\
 r & 0.03 & 66. & 14. & 1.6 & 1.6 \\
 \Delta_{\zeta,{\rm prim}}^2 & 2.41\times 10^9 & 0.64 & 0.64 & 0.25 & 0.25\\
\hline
\end{array}
\end{equation*}
\caption{ Marginalized $1\sigma$ (68\%) errors on parameters for the slow-roll
 inflation model.  The upper table is for the $r_{\rm fid}=0.1$ case; the lower
 table is for the $r_{\rm fid}=0.03$ case.}  \label{table_sr}
\end{figure*}

Using the analytic spectrum from the slow-roll approximation, we
calculate the Fisher matrix and forecast errors on the parameters
attainable form Planck and BBO.  We take ($h, \Omega_b h^2, \Omega_{c}
h^2, \tau, n_S, r, \Delta_{\zeta,{\rm prim}}^2$) as model parameters.
Note that the direct detection is not sensitive to the values of $h,
\Omega_b h^2, \Omega_{c} h^2, \tau$ at all.  Although the transfer
function given in Eq. (\ref{transfer}) includes $h$, $\Omega_m
h^2(=\Omega_b h^2+\Omega_c h^2)$ and $1-\Omega_\Lambda(=\Omega_m)$,
their effects cancel out when the transfer function is evaluated at
higher frequencies ($k\gg k_{\rm eq}$).  So practically we take ($n_S, r,
\Delta_{\zeta,{\rm prim}}^2$) as free parameters in the calculation of
the Fisher matrix for direct detection.  Here, we show two cases of
different fiducial values of $r$: $r_{\rm fid}=0.1$ and $0.03$.  The
fiducial values of the other parameters are taken to be the WMAP5
maximum likelihood.  We use Eq. (\ref{consistencyrelation}) for $n_T$.
The gravitational wave background has a direct detection with SNR=$18.2$
in the case of $r_{\rm fid}=0.1$, and SNR=$8.9$ in $r_{\rm fid}=0.03$.

Figure \ref{fig_sr} shows the constraints in the $n_S-r$ plane expected
from Planck, and those combined with constraints from 10 years of
observation with BBO.  The Planck constraints on $n_S$ and $r$ are not
particularly degenerate as most of the constraint for $r$ comes from the
$B$-mode of the CMB polarization and most of the constraint for $n_S$
comes from the temperature measurements.  The degenerate direction of
direct detection constraints is the direction along which the model
gives the same amplitude of the gravitational wave background spectrum
at the direct detection frequencies.  Since direct detection detects
gravitational waves with a very narrow bandwidth ($0.1-1$Hz), it has
less sensitivity to the tilt of the spectrum by itself and cannot
measure a small deviation from the scale-invariant spectrum.
Considering the fact that direct detection is sensitive only for the
amplitude of the spectrum, the degeneracy line is considered to be the
direction of $\Delta\Omega_{\rm GW}(f=0.2{\rm Hz})=0$.  We can evaluate
the degeneracy line in the $n_S-r$ plane assuming that
$\Delta_{\zeta,{\rm prim}}^2$ is fixed in Eq. (\ref{PTsr}), which yields
$\Delta\Omega_{\rm GW}(k)\propto\Delta(\Delta_{h,{\rm
prim}}^2)=[1-r\ln(k/k_0)/8+r(n_S+r/4-1)\ln^2(k/k_0)/16]\Delta
r+r^2\ln^2(k/k_0)\Delta n_S/16$.  Substituting $k/k_0\simeq 6.5\times
10^{16}$ at $k=2\pi\times 0.2$Hz, we infer that the error ellipse is
elongated along the directions of $0.39\Delta r+0.94\Delta n_S=0$ for
$r_{\rm fid}=0.1$, and $0.77\Delta r+0.084\Delta n_S=0$ for $r_{\rm
fid}=0.03$.  This is consistent with the direction of the main axis of
the error ellipse for Planck+BBO shown in Fig. \ref{fig_sr}.  Note that
the error ellipses obtained by direct detection alone are much more
elongated in these directions, but thanks to the tight constraint on
$n_S$ from CMB, the combined ellipses are less elongated.  Also, the
direct detection constraint itself has no power to distinguish $r$ and
$\Delta_{\zeta,{\rm prim}}^2$ both of which strongly affect the
amplitude of the spectrum [see Eq. (\ref{PTsr})].  However, the CMB
gives a quite tight constraint on $\Delta_{\zeta,{\rm prim}}^2$ and the
degeneracy in the direction of $\Delta_{\zeta,{\rm prim}}^2$ is broken
when they are combined.

As expected, we see direct detection has power mainly in determining the
tensor-to-scalar ratio $r$ to which CMB is not very sensitive.  It is
notable that direct detection tightens the constraint on $r$ more in the
case of $r_{\rm fid}=0.03$ as compared to $r_{\rm fid}=0.1$.  This can
be explained as follows: When $n_S$ and $\Delta_{\zeta,{\rm prim}}^2$ are
fixed, Eq. (\ref{PTsr}) yields $\Delta\Omega_{\rm GW}(k)\propto
[1-r\ln(k/k_0)/8+r(n_S+r/4-1)\ln^2(k/k_0)/16]\Delta r$.  Since the
direct detection does not have sensitivity to the frequency dependence,
it is reasonable to evaluate the uncertainty fixing the frequency at
$0.2$Hz as we did above.  Then the errors on $r$ are expected to be
$\sigma_r\sim\Delta r\propto (1-8.5r+23.4r^2)^{-1}$.  This is an
increasing function when $r\lesssim 0.18$, which means the error becomes
smaller as $r$ decreases.  This happens because the amplitude is not
sensitive to $r$ around $r\sim 0.18$ due to the balance between the
effect of $r$ to increase the amplitude of the tensor-to-scalar ratio
and the effect to decrease the amplitude via the tilt of the spectrum,
$\exp(-r\ln(k/k_0)/8)$, as seen in Eq. (\ref{PTsr}).  Therefore,
$\Omega_{\rm GW}(f=0.2{\rm Hz})$ changes more rapidly with the variation
of $r$ when $r$ is smaller, and this results in the smaller error on $r$
in the $r_{\rm fid}=0.03$ case.  As $r$ decreases more, $\sigma_r$ goes
to a constant value and direct detection no longer has power to
determine $r$.  In that case, BBO may give an upper limit $r\lesssim
0.008$ with a $3\sigma$ confidence level.

The errors on the other parameters are listed as percentages of the
fiducial values in Table \ref{table_sr}.  We see that direct detection
does not help to improve the constraints on CV in this model.

\section{Numerical calculation}
\label{numerical}
In this section, we perform the Fisher matrix calculation using the
spectrum of the gravitational wave background which is obtained by
numerically solving the evolution equation of the gravitational waves.
We calculate the evolution of the scalar field numerically and follow
the evolution of the gravitational waves from the inflation phase up to
the present.  This means the amplitude of the spectrum obtained
numerically reflects the actual Hubble expansion rate when each mode
exits the horizon during inflation.  In contrast, the slow-roll
prediction presented in the previous section may not predict the precise
amplitude of the spectrum at scales far from the CMB scale, since the
spectrum is expressed by making use of a Taylor series approximation.
However, while this numerical approach has the advantage of allowing for
precise evaluation of the amplitude, this numerical approach requires us
to assume an inflation model.  Here, we evaluate errors on each potential
model parameter of quadratic inflation (one potential parameter case),
and natural inflation (two potential parameter case), which can give a
relatively large amplitude of the gravitational wave background.

\subsection{Method}
First, we briefly present the method of our numerical calculation (for
details, see Ref. \cite{Kuroyanagi}).  The evolution equation of the
gravitational wave is simply expressed as 
\begin{equation}
\ddot{h}_\textbf{k}^{\lambda}+3H\dot{h}_\textbf{k}^{\lambda}+\frac{k^2}{a^2}h_\textbf{k}^{\lambda}=0.
\label{heq2}
\end{equation}
We evolve this equation by  calculating $H$ numerically using the following
equations.  During inflation, the evolution of the Hubble expansion is
determined by the scalar field, which decays into radiation in the
reheating phase following inflation.  When considering a case that the
decay rate $\Gamma$ is sufficiently small, the effect of decay can be
simply included into the scalar field equation as \cite{Kolb,Kofman}
\begin{equation}
\ddot{\phi}+(3H+\Gamma)\dot{\phi}+V^{\prime}=0,\label{reheat1}
\end{equation}
and the energy density of the radiation $\rho_r$ generated from the
scalar field obeys the equation,
\begin{equation}
\dot{\rho}_r+4H\rho_r=\Gamma\rho_{\phi}.\label{reheat2}
\end{equation}
Then the Hubble expansion is determined by the energy density of
the $\phi$ field and radiation field,
\begin{equation}
H^2=\frac{8\pi}{3m_{\rm Pl}^2}(\rho_{\phi}+\rho_r),\label{reheat3}
\end{equation}
where the energy density of this scalar field is given as
$\rho_{\phi}=\dot{\phi}^2/2+V$.  After the Universe become well
radiation dominated, we switch to the equation for the Hubble expansion
rate which takes into account the change of $g_*$,
\begin{equation}
H^2=H_0^2\left[\left(\frac{g_*}{g_{*0}}\right)\left(\frac{g_{*s0}}{g_{*s}}\right)^{4/3}\Omega_r
	  a^{-4}+\Omega_m a^{-3}+\Omega_{\Lambda}\right],
\label{Hubble_g}
\end{equation}
where we take $\Omega_r h^2=4.15\times 10^{-5}$ and the other
cosmological parameters are set to be the WMAP maximum likelihood values
given in Sec. \ref{CMBFisher}.

We derive the amplitude of the spectrum by solving the above equations
numerically.  The derivative of the spectrum, which is necessary for the
Fisher calculation, is calculated by performing several calculations in
which we change the parameters slightly.  Here, not only do we take
potential model parameters as parameters for the Fisher matrix, but 
we also take the number of $e$ foldings ${\cal N}$ as a parameter.  The
$e$-folding number is defined as ${\cal N}(k)\equiv\ln(a_{\rm end}/a_k)$,
where $a_{\rm end}$ is the scale factor at the end of inflation and $a_k$ is
the scale factor when the mode exit the horizon ($k=aH$) during
inflation.  When considering a mode which corresponds to the CMB scale
$k_0=0.002{\rm Mpc}^{-1}$, the value of ${\cal N}$ is approximately
given as \cite{Lyth}
\begin{equation}
{\cal N}\simeq 56-\frac{2}{3}\ln\frac{10^{16}{\rm
 GeV}}{\rho_{\rm end}^{1/4}}-\frac{1}{3}\ln\frac{10^9{\rm GeV}}{T_{\rm RH}},
\label{efolds}
\end{equation}
where $\rho_{\rm end}$ is the energy density at the end of inflation and
$T_{\rm RH}$ is the reheating temperature, which directly relates to the
decay rate as \cite{Kolb}
\begin{equation}
T_{\rm RH}\simeq
g_*^{-(1/4)}\left(\frac{45}{8\pi^3}\right)^{1/4}(m_{\rm Pl}\Gamma)^{1/2}.
\label{TRH}
\end{equation}
Note that we do not use Eq. (\ref{efolds}) to obtain the fiducial
value of ${\cal N}$, which is given numerically for a given value of
$\Gamma$.

Here, we have assumed reheating to take place via perturbative decay
\cite{reh1,reh2} as this process can be included simply as in
Eqs. (\ref{reheat1}) and (\ref{reheat2}).  However, in many cases there
can be a stage of preheating (see for example
\citep{koflinstar96,grekof00}) where the inflaton decays via parametric
resonance.  In this paper, we choose a low decay rate consistent with
decay from gravitational effects \cite{nanolisre83} which results in a
reheating temperature of about $10^9$ GeV, and so in the inflation models
we consider there will not be a preheating phase.  This corresponds to
taking $\Gamma\simeq 2$GeV in Eq. (\ref{TRH}).

In order to evaluate the primordial power spectrum parameters, which is
necessary to calculate the Fisher matrix of the CMB, we use the usual slow-roll
formulas to evaluate the parameters given in Sec. \ref{srformula}, by
evaluating them when the mode of $k_{\rm piv}$ exit the horizon $k=aH$ in
the numerical calculation of the background equations.  This method is
reasonable since the slow-roll formulas are sufficiently accurate for
the primordial power spectrum at the length scales probed by the CMB.
In addition to the model parameters taken for the CMB Fisher matrix in the
previous section ($h, \Omega_b h^2, \Omega_{c} h^2, \tau, n_S, r,
\Delta_{\zeta,{\rm prim}}^2$), we also include the running of the scalar
spectrum index $\alpha_S$.

\subsection{Quadratic Inflation}

\begin{figure*}[htbp]
 \includegraphics[width=0.48\textwidth]{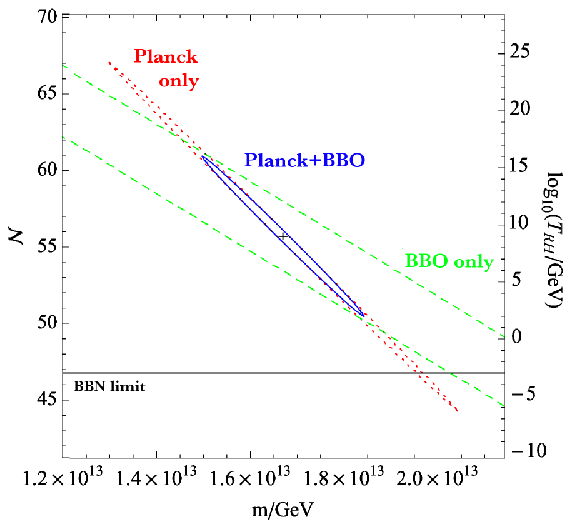} \caption{ Constraint
 on parameters of the quadratic potential.  The dotted line (red) and
 the dashed line (green) show marginalized $2\sigma$ confidence
 region in the $m-{\cal N}, \log_{10} T_{\rm RH}$ plane allowed by
 Planck and BBO, respectively, and the solid line (blue) shows the
 combined constraint. The black horizontal line shows the BBN limit on
 reheating temperature, $T_{\rm RH}>1$MeV. }  \label{fig_quadratic}

\makeatletter
\def\@captype{table}
\makeatother

\begin{equation*}
\begin{array}{c|c|cccc}
\hline
 \text{Variable} & \text{Fiducial value} & \text{$\%$ Error Planck only}
 & \text{$\%$ Error Planck+BBO} & \text{$\%$ Error CV} & \text{$\%$ Error CV+BBO} \\
\hline
 h & 0.724 & 0.64 & 0.49 & 0.15 & 0.15 \\
 \Omega_c h^2 & 0.108 & 0.84 & 0.62 & 0.2 & 0.2 \\
 \Omega_b h^2 & 0.227 & 0.52 & 0.5 & 0.17 & 0.17 \\
 \tau & 0.089 & 3.9 & 3.7 & 2.4 & 2.4 \\
 m/{\rm GeV} & 1.66\times 10^{13} & 8.8 & 4.2 & 1.2 & 1.1 \\
 {\cal N} & 55.7 & 8.3 & 3.9 & 1.1 & 1.1 \\
 \log_{10} (T_{\rm RH}/{\rm GeV}) & 9.0 & 67. & 31. & 9.1 & 8.8\\
\hline
\end{array}
\end{equation*}
\caption{Marginalized $1\sigma$ errors on parameters for the quadratic
inflation model.}\label{table_quadratic}
\end{figure*}

First, we investigate the case of the quadratic potential,
\begin{equation}
V=\frac{1}{2}m^2\phi^2.
\end{equation}
We take two model parameters: the mass of the scalar field $m$ and
$e$-folds number ${\cal N}$, or reheating temperature $\log_{10} (T_{\rm
RH}/{\rm GeV})$.  The error in ${\cal N}$ can be converted to the error
in $\log_{10} (T_{\rm RH}/{\rm GeV})$ by Eq. (\ref{efolds}) as
$\sigma_{\cal N}=0.77\sigma_{\log_{10} (T_{\rm RH}/{\rm GeV}))}$.  The
fiducial value of $m$ is determined to satisfy the normalization of the
scalar perturbations $\Delta_{\zeta,{\rm prim}}^2=2.41\times 10^{-9}$,
and ${\cal N}$ is determined by $T_{\rm RH}=10^9$GeV.  They are derived
numerically as $(m,{\cal N})=(1.66\times 10^{13}{\rm GeV}, 55.7)$.  In
this case, the gravitational wave background is detected with SNR
$=16.5$ by the direct detection experiment.

Figure \ref{fig_quadratic} shows the confidence contours in the $m - {\cal
N}$ plane, expected from Planck, BBO and both combined.  We see Planck
gives good constraints on the mass of the scalar field $m$, but has less
power to determine the $e$-folds number ${\cal N}$.  Although the BBO
constraint is weaker than Planck, it can break the strong degeneracy in
the parameters since the degenerate directions are slightly different,
and improve the errors on both of the parameters.  Table
\ref{table_quadratic} shows how much the errors from Planck and CV
decrease when they are combined with the constraints from BBO.

Reminding the reader that direct detection cannot distinguish models
which give the same amplitude at the direct detection frequencies, the
degenerate direction is considered to be the direction in which
$\Omega_{\rm GW}\propto\Delta_{h, {\rm prim}}^2\propto H^2|_{k=aH}$ is
constant [see Eq. (\ref{PTprim})].  Using the relations that the Hubble
expansion rate during inflation is given as $H(k)^2\propto
V(k)=m^2\phi(k)^2/2$, and $\phi(k)$ relates to the $e$-folding number as
$\phi(k)^2=2{\cal N}(k)+1$ in the case of the quadratic potential model,
the parameters give the same spectrum amplitude in the direction of
$m^2[2{\cal N}(k)+1]=const$.  Therefore, the degenerate line is
considered to be $\Delta m/m+\Delta {\cal N}/[2{\cal N}(k)+1]=0$.
Substituting ${\cal N}(k=2\pi\times 0.2{\rm Hz})\simeq 16.4$ which
corresponds to the direct detection scale, the degeneracy direction of
the direct detection constraint is estimated as $\Delta m/(1.66\times
10^{13}{\rm GeV})+\Delta {\cal N}/33.7\simeq 0$, which is consistent
with the result shown in Fig. \ref{fig_quadratic}.

We also show the constraints in terms of $\log_{10}(T_{\rm RH}/{\rm
GeV})$ instead of ${\cal N}$.  It is notable that the marginalized error
on the reheating temperature is $\sigma_{\log_{10}(T_{\rm RH}/{\rm
GeV})}\sim 6.0$ by Planck alone, which means the BBN lower limit of
$\log_{10}(T_{\rm RH}/{\rm GeV})=-3$ is less than $2\sigma$ away from
the fiducial value without the direct detection constraint.  When it is
combined with direct detection constraint, the error is reduced to be 
$\sigma_{\log_{10}(T_{\rm RH}/{\rm GeV})}\sim 2.8$, and the BBN limit is
ruled out at more than $4\sigma$.

\subsection{Natural inflation}

\begin{figure*}[htbp]
 \includegraphics[width=0.48\textwidth]{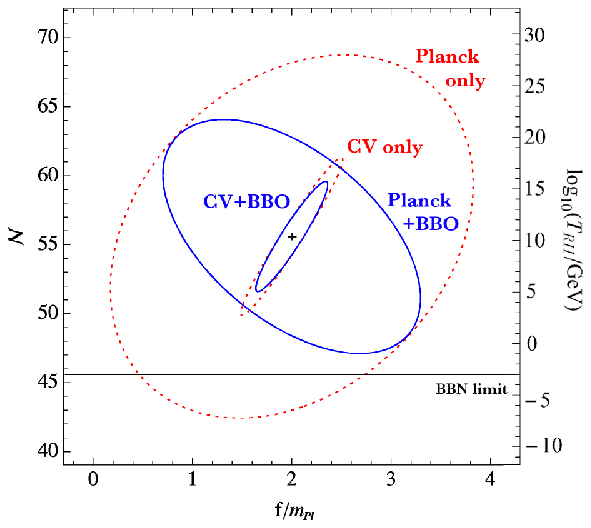} \hfill
 \includegraphics[width=0.48\textwidth]{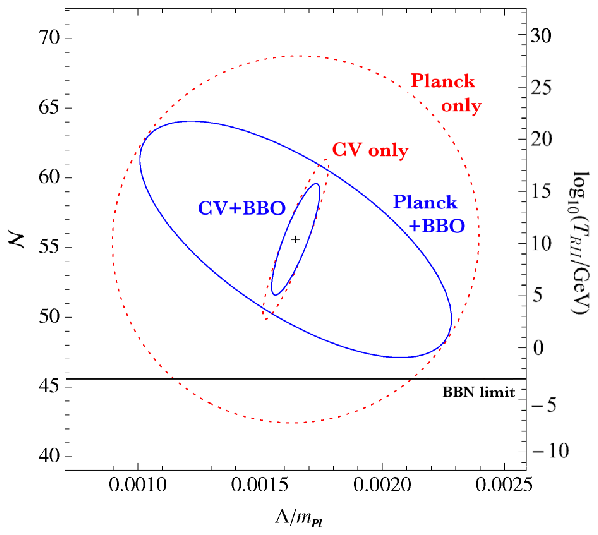} \caption{
 Forecasted marginalized $2\sigma$ constraints on parameters
 for the natural inflation model with $f=2m_{\rm Pl}$.  CMB (red dotted) and
 CMB plus direct detection (blue solid) are shown in the
 $f/m_{\rm Pl}-{\cal N}$ and $\Lambda/m_{\rm Pl}-{\cal N}$
 plane, respectively. In each case the larger contour is when the CMB is
 Planck and the smaller is when the CMB is cosmic variance limited.  }
 \label{fig_natural} \makeatletter \def\@captype{table} \makeatother

\begin{equation*}
\begin{array}{c|c|cccc}
 \hline
 \text{Variable} & \text{Fiducial value} & \text{$\%$ Error Planck only}
 & \text{$\%$ Error Planck+BBO} & \text{$\%$ Error CV} & \text{$\%$ Error CV+BBO} \\
 \hline
 h & 0.724 & 0.7 & 0.66 & 0.15 & 0.15 \\
 \Omega_c h^2 & 0.108 & 0.92 & 0.86 & 0.21 & 0.21 \\
 \Omega_b h^2 & 0.227 & 0.54 & 0.53 & 0.18 & 0.18 \\
 \tau & 0.089 & 3.7 & 3.7 & 2.4 & 2.4 \\
 \Lambda/m_{\rm Pl} & 1.07\times 10^{\text{-3}} & 8.8 & 7.5 & 0.57 & 0.56 \\
 f/m_{\rm Pl} & 1.0 & 5.5 & 4.3 & 2.1 & 2. \\
 {\cal N} & 55.2 & 11. & 10. & 2.3 & 2.2 \\
 \log_{10} (T_{\rm RH}/{\rm GeV}) & 9.0 & 85. & 83. & 18. & 18.\\
 \hline
\end{array}
\end{equation*}

\begin{equation*}
\begin{array}{c|c|cccc}
 \hline
 \text{Variable} & \text{Fiducial value} & \text{$\%$ Error Planck only} & \text{$\%$
  Error Planck+BBO} & \text{$\%$ Error CV} & \text{$\%$ Error CV+BBO} \\
 \hline
 h & 0.724 & 0.69 & 0.52 & 0.15 & 0.15 \\
 \Omega_c h^2 & 0.108 & 0.91 & 0.66 & 0.22 & 0.21 \\
 \Omega_b h^2 & 0.227 & 0.53 & 0.5 & 0.19 & 0.18 \\
 \tau & 0.089 & 3.9 & 3.8 & 2.6 & 2.4 \\
 \Lambda/m_{\rm Pl} & 1.64\times 10^{\text{-3}} & 18. & 16. & 3.3 & 2.4 \\
 f/m_{\rm Pl} & 2.0 & 37. & 26. & 10. & 7.3 \\
 {\cal N} & 55.6 & 9.5 & 6.1 & 4.1 & 2.9 \\
 \log_{10} (T_{\rm RH}/{\rm GeV}) & 9.0 & 77. & 49. & 33. & 23.\\
 \hline
\end{array}
\end{equation*}
\caption{Marginalized $1\sigma$ errors on parameters for the natural
 inflation model.  The upper table is for the $f=m_{\rm Pl}$ case; the lower
 table is for the $f=2m_{\rm Pl}$ case.}\label{table_natural}
\end{figure*}

Next, we investigate natural inflation for  which the  potential is given as
\cite{natural1,natural2,natural3}
\begin{equation}
V=\Lambda^4\left[1\pm\cos\left(\frac{N\phi}{f}\right)\right].
\end{equation}
Here, we set $N=1$ and take the positive sign.  In this case, we take
three variables, $\Lambda/m_{\rm Pl}$, $f/m_{\rm Pl}$ and ${\cal N}$, to be
model parameters.  We investigate two cases of different fiducial values
of $f$, which are taken to be $f=m_{\rm Pl}$ and $f=2m_{\rm Pl}$, and the
fiducial value of $\Lambda$ is determined by the normalization of
$\Delta_{\zeta,{\rm prim}}^2$.  The potential becomes more quadratic and
predicts larger amplitudes of the gravitational wave background as $f$
increases.  The fiducial values are, respectively, $(\Lambda/m_{\rm Pl},
f/m_{\rm Pl}, {\cal N})=(1.07\times 10^{-3}, 1.0, 55.2)$ for $f=m_{\rm Pl}$ and
$(1.64\times 10^{-3}, 2.0, 55.6)$ for $f=2m_{\rm Pl}$.  The gravitational
wave background is detected with SNR $=8.4$ in the case of $f=m_{\rm Pl}$,
and SNR $=14.9$ in the case of $f=2m_{\rm Pl}$.

Figure \ref{fig_natural} shows the error ellipsoids in the $f=2m_{\rm Pl}$
case.  Table \ref{table_natural} gives errors on each parameters.  The
$f=2m_{\rm Pl}$ case is essentially the quadratic case and so the errors on
${\cal N}$ are about the same.  A Taylor expansion around the bottom of
the potential gives the effective mass to be $m\approx\Lambda^2/f$ when
$f\gg1$. Therefore the $\Lambda$ and $f$ parameters are expected to
become correlated in the $f=2m_{\rm Pl}$ case and less so in the $f=m_{\rm Pl}$
case.  This is consistent with the errors of $f$ and $\Lambda$ being
larger in the $f=2m_{\rm Pl}$ case.  The gravitational waves are too low for
the direct detection experiment to have much effect in the $f=m_{\rm Pl}$
case.  The error in the reheating temperature is quite large.  Even in
the $f=2m_{\rm Pl}$ case the Planck and direct detection data still has a
fiducial value $T_{\rm RH}=10^9$GeV less than $3\sigma$ away from the
BBN limit of $T_{\rm RH}=10^{-3}$GeV.  Unlike in the quadratic inflation
case, direct detection is still useful to tighten the constraints from
CV for the $f=2m_{\rm Pl}$ model.

\section{Conclusion}
\label{conclusion}
This paper is aimed at studying how direct detection of the inflationary
gravitational wave background will determine inflationary parameters and
how it will complement  future CMB polarization experiments.  An
attractive feature of these two different methods of observation is that
they probe two different frequencies and provide independent
information.  By calculating the Fisher matrix, we have investigated the
degree to which the errors on model parameters obtained from CMB
experiments will be reduced by direct detection of the gravitational
waves in the BBO experiment.  We have presented two different types of
methods to calculate the Fisher matrix: One is evaluated analytically by
making use of the slow-roll approximation, and the other is evaluated
numerically for the sake of more accurate predictions.  In the second
case, we also allow the temperature of reheating to be a free parameter.

In both cases, we have shown that the two different observations have
different directions of parameter degeneracy and that this degeneracy is
broken when they are combined.  The degeneracy of a direct detection
experiment is directed to the direction in which the model parameters
give the same amplitude at the direct detection frequencies.  Although
our result indicates that the BBO experiment provides a larger error in
parameter estimation than CMB experiments, it most certainly has the
power to tighten constraints from Planck.  We also found that for
natural inflation, direct detection could even help to improve a cosmic
variance-limited CMB experiment. It would be interesting to check what
the improvements could be in multifield inflation models where the
number of parameters would be greater.

Constraints on parameters which are defined by the slow-roll parameters
were calculated by using an analytic spectrum of the gravitational wave
background.  BBO direct detection worked to tighten the constraint from
Planck mainly on the tensor-to-scalar ratio.  On the other hand, we have
confirmed the intuitive fact that it does not have the power to constrain
the other cosmological parameters, which makes sense since the
tensor-to-scalar ratio is the major parameter in determining the
amplitude of the gravitational wave background.  Constraints on
inflation potential parameters have been evaluated by using an accurate
amplitude of the spectrum obtained from numerical calculations, for both
the quadratic inflation and natural inflation cases.  In addition to
potential parameters, we also have taken the $e$-folds number, which
corresponds to the reheating temperature, as a parameter and have found that
BBO direct detection has power to tighten constraints on both of the
parameters.  For quadratic inflation, we found without BBO direct detection,
Planck could only rule out a BBN reheating temperature at the $2\sigma$
level.  However with BBO direct detection it could rule this  out at almost
the $4\sigma$ level. In the case of natural inflation (with
$f=2m_{\rm Pl}$), we found that a cosmic variance-limited CMB experiment
could only rule out a BBN reheating temperature at the $4\sigma$ level
while with BBO direct detection it could be ruled out at almost the
$6\sigma$ level.

Note that, when the signal to noise ratio is low, the Fisher matrix can
underestimate the error bars (see for example \cite{Vallisneri}).
Therefore our Planck only results may be overly optimistic, but the
combined Planck plus BBO and the cosmic variance CMB results should be
more accurate.  In future work, we plan to redo this analysis using a
Markov Chain Monte Carlo method, which will also make it easier to add
more detailed priors on the reheating temperature.

For reference, let us mention another project aimed at
detecting the inflationary gravitational wave background directly,
called DECIGO
\cite{decigo}.  It has similar specifications to BBO; the target
frequency is almost the same, but the sensitivity is a little smaller
than that of BBO.  Because of the fact that the direction of the parameter
degeneracy is determined only by the amplitude of the gravitational wave
at direct the detection scale, the degenerate direction of constraints
from DECIGO is the same as the one from BBO.  However, since DECIGO has
less sensitivity, the size of the error ellipse is considered to be
bigger than BBO and the Fisher matrix analysis may not be applicable
unless the tensor-to-scalar ratio is relatively large.  (For example,
while BBO detects the gravitational wave background of $r=0.1$ with
SNR$=18.2$, DECIGO detects this with SNR$=4.1$, which may be out of the
validity of the Fisher matrix analysis.)  For this reason, we have
presented the results only for BBO in this paper.

\section*{Acknowledgments}
The authors are grateful to Kavilan Moodley and Naoki Seto for useful
comments.  SK would like to thank Tsutomu Takeuchi for his help with the
study on the Fisher matrix analysis.  This research is supported by
Grant-in-Aid for Nagoya University Global COE Program, "Quest for
Fundamental Principles in the Universe: From Particles to the Solar
System and the Cosmos."  CG is supported by the Beecroft Institute for
Particle Astrophysics and Cosmology.

\appendix
\section{Suppression factor for the cosmological constant}
\label{omegalambda}
Reference \cite{Zhang} shows that the current acceleration of the Universe
suppresses the amplitude of the inflation-produced gravitational waves by
a factor of $\Omega_m/\Omega_{\Lambda}$.  This is derived using the fact
that the scale factor when the mode enter the horizon during the
accelerating stage satisfies the relation $a_{\rm hc}\propto k$.
However, it is not appropriate to apply this to the case of our
Universe, $\Omega_{\Lambda}=0.7,\Omega_{m}=0.3$, which just starts to
enter the cosmological constant-dominated Universe and is still not in
the middle of a de Sitter phase of exponential expansion.

Here, we propose a new suppression factor which gives a better
approximation.  From the behavior of the inflationary gravitational waves
that $h_{\bf k}$ keeps constant outside the horizon and decreases
proportional to $a^{-1}$ after entering the horizon, the transfer
function is considered to be written as $T_h(k)=|h_{{\bf k},0}|/|h_{{\bf
k},{\rm hc}}|=a_{\rm hc}/a_0$.  The suppression factor is measured by comparing
the transfer functions in the case of the Universe without the
cosmological constant, which we label with a subscript $1$, and in the
case with the cosmological constant, which we label with a subscript
$2$.  Therefore, the value of  interest, how much the amplitude of
the gravitational waves is suppressed by the cosmological constant, is
\begin{equation}
\frac{T_{h,2}(k)}{T_{h,1}(k)}=\frac{a_{{\rm hc},2}}{a_{{\rm hc},1}}.
\label{sf1}
\end{equation}
We set the Hubble parameter $H_0$ and the density parameter of radiation
$\Omega_r$ to be the same value in both cases.  The only difference is
the existence of the cosmological constant, $\Omega_{\Lambda}$.  If we
assume a flat Universe, then the density parameter of matter is described
with the amount of the cosmological constant as
$\Omega_{m,2}=1-\Omega_{r}-\Omega_{\Lambda}$, while
$\Omega_{m,1}=1-\Omega_{r}$ if there is no cosmological constant.  With
using the relation $k=a_{\rm hc}H_{\rm hc}$ and rewriting the Hubble
parameter in terms of the cosmological parameters, Eq. (\ref{sf1})
becomes
\begin{eqnarray}
&&\frac{a_{{\rm hc},2}}{a_{{\rm hc},1}}
=\frac{k/H_{{\rm hc},2}}{k/H_{{\rm hc},1}}\nonumber\\
&&=\frac{H_0\sqrt{(1-\Omega_{r})a_{{\rm hc},1}^{-3}+\Omega_{r}a_{
 {\rm hc},1}^{-4}}}{H_0\sqrt{(1-\Omega_{r}-\Omega_{\Lambda})a_{
 {\rm hc},2}^{-3}+\Omega_{r}a_{{\rm hc},2}^{-4}+\Omega_{\Lambda}}}.
\label{App2}
\end{eqnarray}
Let us consider a mode which enters the horizon during the
matter-dominated phase.  Since the contribution of radiation and the
cosmological constant terms to the Hubble expansion is negligible during
this phase, Eq. (\ref{App2}) becomes
\begin{equation}
\frac{a_{{\rm hc},2}}{a_{{\rm hc},1}}
\simeq\sqrt{\frac{(1-\Omega_{r})}{(1-\Omega_{r}-\Omega_{\Lambda})}\left(\frac{a_{{\rm hc},1}}{a_{{\rm hc},2}}\right)^{-3}}.
\end{equation}
Neglecting the radiation density parameter, which is much smaller than
1, we find the suppression factor is approximately
\begin{equation}
\frac{a_{{\rm hc},2}}{a_{{\rm hc},1}}
\simeq 1-\Omega_{\Lambda}.
\end{equation}



\begin{thebibliography}{99}

\bibitem{inf1} A. H. Guth, Phys. Rev. D {\bf 23}, 347 (1981).

\bibitem{inf2} A. Albrecht and P. J. Steinhardt,
  Phys. Rev. Lett. {\bf 48}, 1220 (1982).

\bibitem{inf3}  A. Linde, Phys. Lett. B {\bf 108}, 389 (1982).

\bibitem{grav1} B. Allen, Phys. Rev. D {\bf 37}, 2078 (1988).

\bibitem{grav2} V. Sahni, Phys. Rev. D {\bf 42}, 453 (1990).

\bibitem{grav3} L. P. Grishchuk and Y. V. Sidorov, Phys. Rev. D {\bf
  42}, 3413 (1990); L. P. Grishchuk and M. Solokhin, Phys. Rev. D {\bf
  43}, 2566 (1991).

\bibitem{CMBp1} U. Seljak and M. Zaldarriaga, Phys. Rev. Lett. {\bf
  78}, 2054 (1997); M. Zaldarriaga and U. Seljak, Phys. Rev. D {\bf
  55}, 1830 (1997).

\bibitem{CMBp2} M. Kamionkowski, A. Kosowsky, and A. Stebbins,
  Phys. Rev. Lett. {\bf 78}, 2058 (1997); Phys. Rev. D {\bf 55}, 7368
  (1997).

\bibitem{Planck} http://www.rssd.esa.int/index.php?project=PLANCK.

\bibitem{decigo} N. Seto, S. Kawamura, and T. Nakamura,
  Phys. Rev. Lett. {\bf 87}, 221103 (2001); S. Kawamura {\it et al.},
  Class. Quant. Grav. {\bf 23}, S125 (2006).

\bibitem{bbo} E. S. Phinney {\it et al.}, "The Big Bang Observer", NASA
Mission Concept Study (2003); G. M. Harry,
  P. Fritschel, D. A. Shaddock, W. Folkner and E. S. Phinney,
  Class. Quant. Grav. {\bf 23}, 4887 (2006) [Erratum-ibid. {\bf 23},
    7361 (2006)].

\bibitem{Smith1} T. L. Smith, M. Kamionkowski and A. Cooray,
  Phys. Rev. D {\bf 73}, 023504 (2006); Phys. Rev. D {\bf 78}, 083525
	(2008).

\bibitem{Smith2} T. L. Smith, H. V. Peiris, and A. Cooray,
	Phys. Rev. D {\bf 73}, 123503 (2006).

\bibitem{Ungarelli} C. Ungarelli {\it et al.}, Classical Quantum
	Gravity {\bf 22}, 955 (2005).

\bibitem{Lewis} A. Lewis, A. Challinor, and A. Lasenby,
	Astrophys. J. {\bf 538}, 473 (2000).

\bibitem{Kuroyanagi} S. Kuroyanagi, T. Chiba and N. Sugiyama,
	Phys. Rev. D {\bf 79}, 103501 (2009).

\bibitem[Zaldarriaga et.al.(1997)]{zalspesel97} M. Zaldarriaga,
	D. Spergel, U. Seljak, Astrophys. J. {\bf 488} 1 (1997).

\bibitem[Bond et. al.(1997)]{bonefsteg97} J. R. Bond, G. Efstathiou and
	M. Tegmark, Mon. Not. R. Astron. Soc. {\bf 291}, L33 (1997).

\bibitem[Efstathiou et.al.(2009)]{efgrapac09} G. Efstathiou, S. Gratton
	and F. Paci, Mon. Not. R. Astron. Soc. {\bf 397}, 1355 (2009).

\bibitem{Baumann} D. Baumann {\it et al.}, arXiv:0811.3919
	[astro-ph].

\bibitem{WMAP5yr} J. Dunkley {\it et al.}, Astrophys. J. Suppl. {\bf
	180} 306 (2009).

\bibitem{Michelson} P. F. Michelson, Mon. Not. R. Astron. Soc. {\bf
	227}, 933 (1987).

\bibitem{Christensen} N. Christensen, Phys. Rev. D {\bf 46}, 5250
	(1992).

\bibitem{Flanagan} E. E. Flanagan, Phys. Rev. D {\bf 48}, 2389 (1993).

\bibitem{Allen} B. Allen and J. D. Romano, Phys. Rev. D {\bf 59}, 102001
	(1999).

\bibitem{Seto} N. Seto, Phys. Rev. D {\bf 73}, 063001 (2006).

\bibitem{Prince} T. A. Prince, M. Tinto, S. L. Larson, and
	J. W. Armstrong, Phys. Rev. D {\bf 66}, 122002 (2002).

\bibitem{LISA} P. Bender {\it et al.}, "LISA Pre-Phase A Report", second
	Edition, MPQ 233, (1998);
	http://lisa.gsfc.nasa.gov/Documentation/ppa2.08.pdf

\bibitem{Cutler} C. Cutler and J. Harms, Phys. Rev. D 73, 042001 (2006).

\bibitem{Cornish1} N. J. Cornish and S. L. Larson, Classical Quantum
	Gravity {\bf 18}, 3473 (2001).

\bibitem{Cornish2} N. J. Cornish, Phys. Rev. D {\bf 65}, 022004 (2001).

\bibitem{Corbin} V. Corbin and N. Cornish,
	Classical Quantum Gravity {\bf 23}, 2435 (2006).

\bibitem{Lidsey} J. E. Lidsey {\it et al.}, Rev. Mod. Phys. {\bf 69},
  373 (1997).

\bibitem{Kosowsky} A. Kosowsky and M. S. Turner, Phys. Rev. D {\bf
  52}, R1739 (1995).

\bibitem{Turner} M. S. Turner, M. White and J. E. Lidsey, Phys. Rev. D
	{\bf 48}, 4613 (1993); M. S. Turner, Phys. Rev. D {\bf 55}, R435
	(1997).

\bibitem{Zhang} Y. Zhang, Y. Yuan, W. Zhao and Y. T. Chen, Classical
	Quantum Gravity {\bf 22}, 1383 (2005).

\bibitem{Schwarz}
 D. J. Schwarz, Mod. Phys. Lett. A {\bf 13}, 2771 (1998).

\bibitem{Watanabe} Y. Watanabe and E. Komatsu, Phys. Rev. D {\bf 73},
  123515 (2006).

\bibitem{Nakayama} K. Nakayama, S. Saito, Y. Suwa and J. Yokoyama,
	Phys. Rev. D {\bf 77}, 124001 (2008);
	J. Cosmol. Astropart. Phys. 06 (2008) 020.

\bibitem{Kolb} E. W. Kolb and M. S. Turner, {\it The Early Universe}
  (Westview Press, Boulder, CO, 1990).

\bibitem{Kofman} L. Kofman, A. Linde and A. A. Starobinsky,
  Phys. Rev. D {\bf 56}, 3258 (1997).

\bibitem{Lyth} D. H. Lyth and A. R. Liddle, {\it The Primordial Density
	Perturbation} (Cambridge University Press, 2009).

\bibitem{reh1} A. D. Dolgov and A. D. Linde, Phys. Lett. B {\bf
  116}, 329 (1982).

\bibitem{reh2} L. F. Abbott, E. Farhi and M. B. Wise, Phys. Lett. B
  {\bf 117}, 29 (1982).

\bibitem{koflinstar96}
  L. Kofman, A. D. Linde and A. A. Starobinsky,
  Phys. Rev. Lett. {\bf 76}, 1011 (1996).
  
\bibitem{grekof00} P. B. Greene and L. Kofman,
  Phys. Rev.  D {\bf 62}, 123516 (2000).

\bibitem{nanolisre83}
  D. V. Nanopoulos, K. A. Olive and M. Srednicki,
  Phys. Lett.  B {\bf 127}, 30 (1983).

\bibitem{natural1} K. Freese, J. A. Frieman and A. V. Olinto,
	Phys. Rev. Lett. {\bf 65}, 3233 (1990).

\bibitem{natural2}  F. C. Adams, J. R. Bond, K. Freese, J. A. Frieman and
	A. V. Olinto, Phys. Rev. D {\bf 47}, 426 (1993).

\bibitem{natural3} C. Savage, K. Freese and W. H. Kinney, Phys. Rev. D
	{\bf 74}, 123511 (2006).

\bibitem{Vallisneri} M. Vallisneri, Phys. Rev. D {\bf 77}, 042001
	(2008).

\end{thebibliography}
\end{document}